\documentclass{article}
\usepackage[top=1in, bottom=1in, left=1in, right=1in]{geometry}
\usepackage[utf8]{inputenc}
\usepackage{etoolbox}
\usepackage{xcolor}
\usepackage{graphicx}
\usepackage[permil]{overpic}
\usepackage{cprotect}
\usepackage{amsmath}
\graphicspath{ {./figures/} }
\usepackage{tikz}
\usetikzlibrary{shapes, arrows.meta, positioning, chains}
\usepackage{adjustbox}
\usepackage{hyperref}
\usepackage{xcolor}
\usepackage{enumerate}
\usepackage{amsfonts}
\usepackage[normalem]{ulem}
\usepackage[numbers]{natbib}
\usepackage{algorithm}
\usepackage{algpseudocode}
\usepackage{subcaption}
\usepackage{amssymb}
\usepackage{multirow}

\title{Coarse graining and reduced order models for plume ejection dynamics}
\author{Ike Griss Salas$^*$, Megan R. Ebers$^*$, Jake Stevens-Haas$^*$, and J. Nathan Kutz$^{*,\ddag}$\\
\small $^*$ Department of Applied Mathematics, University of Washington, Seattle, WA 98195 USA\\
\small $^\ddag$ Department of Electrical and Computer Engineering, University of Washington, Seattle, WA 98195}

\begin{document}

\maketitle

\begin{abstract}
Monitoring the atmospheric dispersion of pollutants is increasingly critical for environmental impact assessments. High-fidelity computational models are often employed to simulate plume dynamics, guiding decision-making and prioritizing resource deployment. However, such models can be prohibitively expensive to simulate, as they require resolving turbulent flows at fine spatial and temporal resolutions.
Moreover, there are at least two distinct dynamical regimes of interest in the plume:  (i) the initial ejection of the plume where turbulent mixing is generated by the shear-driven Kelvin-Helmholtz instability, and (ii) the ensuing turbulent diﬀusion and advection which is often modeled by the Gaussian plume model. 
We address the challenge of modeling the initial plume generation. Specifically, we propose a data-driven framework that identifies a reduced-order analytical model for plume dynamics -- directly from video data. We extract a time series of plume center and edge points from video snapshots and evaluate different regressions based to their extrapolation performance to generate a time series of coefficients that characterize the plume's overall direction and spread. We regress to a sinusoidal model inspired by the Kelvin-Helmholtz instability for the edge points in order to identify the plume's dispersion and vorticity. Overall, this reduced-order modeling framework provides a data-driven and lightweight approach to capture the dominant features of the initial nonlinear point-source plume dynamics, agnostic to plume type and starting only from video. The resulting model is a pre-cursor to standard models such as the Gaussian plume model and has the potential to enable rapid assessment and evaluation of critical environmental hazards, such as methane leaks, chemical spills, and pollutant dispersal from smokestacks.

\end{abstract}

\section{Introduction}

Wildfires, volcanic eruptions, chemical spills, and industrial emissions disperse plumes of pollutants into the atmosphere. For example, approximately 1.2 to 2.6 million tons of methane leakage occurs from natural gas pipelines per year \citep{ mcvay2023methane}. Locating and monitoring air pollutant dispersion like methane leaks is an important action towards reducing harmful climate pollution \cite{ocko2021acting, myhre2014anthropogenic, ocko2018rapid}, it is also important issue in national security~\cite{settles2006fluid}.

By modeling plumes, scientists, public health officials, and emergency responders can understand the concentration and spread of pollutants carried by the plume to evaluate health, safety, and environmental risks.  This paper seeks to develop a data-driven reduced order model for the initial plume formation which is driven by the Kelvin-Helmholtz instability.  This is critical part of the overall plume dynamics and is a potential pre-cursor to more standard models for the ensuing diﬀusion and advection dynamics of the plume,  thus scientists and emergency responders can use such reduced models to understand the overall drift and dispersion of pollutants.  While other computational models to model plumes exist, the aim for our model is to construct models directly from data which characterizes the initial plume formation; the only required data collection being video such as from a smartphone. \\

Mathematical modeling of plume dynamics has allowed the prediction, management, and understanding of air pollutant dispersion \cite{briggs1975plume}. Mathematical models can be used inversely for management, such as source localization or emission rate estimation; forward models can be used for prediction, like emergency response planning or designing emission control strategies. A standard mathematical model used in the community is known as the Gaussian plume model~\cite{arya1999air,stockie2011mathematics} which assumes the plume to be a Gaussian spreading through a coarse-grained description of the turbulent diﬀusion and advection of the atmosphere.  The Gaussian plume model focuses on the dynamics once plume has already formed.  A spectrum of additional dispersion models exist, with the most powerful being computational fluid dynamics (CFD) models, which offer high-fidelity simulations of complex flow scenarios \cite{tominaga2018cfd, flaherty2007computational, ma2022cfd, cloete2009cfd, hanna2006detailed}. However, CFD simulations of plume dynamics can be impractical for rapid assessment and evaluation in time-critical scenarios. \\

While high-fidelity simulations represent the state of the art in accurately tracking plume dispersion and localizing sources over time, their computational demands are often prohibitively high. Early research into air pollution dispersion dates back nearly 100 years, with Bosanquet and Pearson evaluating the spread of chimney smoke \cite{bosanquet1936spread}. In the following decade, mathematical models evolved to incorporate additional complexities in air pollution dispersion equations, such as vertical dispersion, crosswind dispersion, and ground effects \cite{sutton1947problem}.\\

In the 1960s, environmental control regulations in the United States triggered a surge in computational approaches to modeling pollutant plumes \cite{irwin2002historical}. The foundation for these computational approaches was the Gaussian dispersion model of continuous and buoyant air pollution plumes \cite{beychok2005fundamentals}. Additionally, the Briggs equations enabled trajectory modeling of more complex plume behaviors \cite{briggs1965plume}.

Advancements in computational technology have since enabled high-fidelity CFD simulations to more accurately model particulate dispersion in complex environments and flow regimes \cite{joseph2020reconciling, di2007simulations, silvester2009computational, lowndes2008application}. However, such simulations can easily become computationally intractable for large spatial domains and fine spatial resolutions \cite{hanna2006detailed}. This is further compounded by the complexity of such simulations and their sensitivity to model parameters \cite{pan2013enhanced, cloete2009cfd}. Accurately modeling plume dynamics is highly dependent on the specific type of plume, such as methane or hydrothermal. This dependency limits the generalization of a model without significant upfront effort to fine-tune it using first principles.

To address these challenges, reduced order models (ROMs) and data-driven modeling techniques have emerged as practical solutions.  \\

Reduced order models (ROMs) can significantly reduce computational requirements while preserving accurate approximations of complex phenomena. Classical approaches like proper orthogonal decomposition and dynamic mode decomposition \cite{lumley1967structure, holmes2012turbulence, rowley2005model, kutz2016dynamic, tu2013dynamic} are widely used to find a low-dimensional and data-driven basis that capture dominant flow structure. These methods have been used to study urban airflow and pollutant dispersion \cite{masoumi2022review, xiao2019machine, nony2023reduced}. However, two major drawbacks of these methods are: (i) they have limited spatial extrapolation ability, and (ii) they learn optimal linear subspaces, while fluid dynamics such as plume behavior are strongly nonlinear. Indeed, nonlinear dimensionality reduction techniques have demonstrated improved system characterization of fluid flow \cite{fu2023non, fresca2021comprehensive, xiang2021non}. However, the black-box nature of deep learning paradigms like autoencoder architectures can limit physical interpretability. \\

\begin{figure}[htbp]
    \centering
    \begin{subfigure}{0.4\linewidth}
        \centering
        \includegraphics[width=\linewidth]{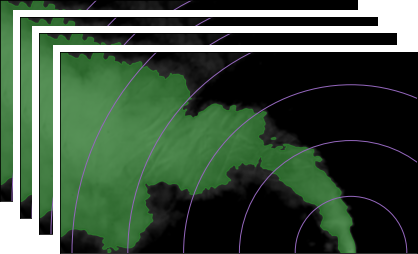}
        \caption{}
    \end{subfigure}
    \begin{subfigure}{0.3\linewidth}
        \centering
        \includegraphics[width=\linewidth]{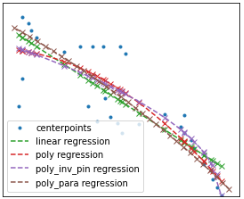}
        \caption{}
    \end{subfigure}

    \vspace{0.5cm}

    \begin{subfigure}{0.3\linewidth}
        \centering
        \includegraphics[width=\linewidth]{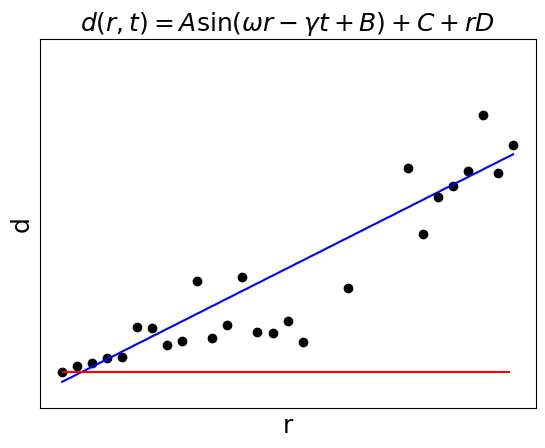}
        \caption{}
    \end{subfigure}
    \begin{subfigure}{0.4\linewidth}
        \centering
        \includegraphics[width=\linewidth]{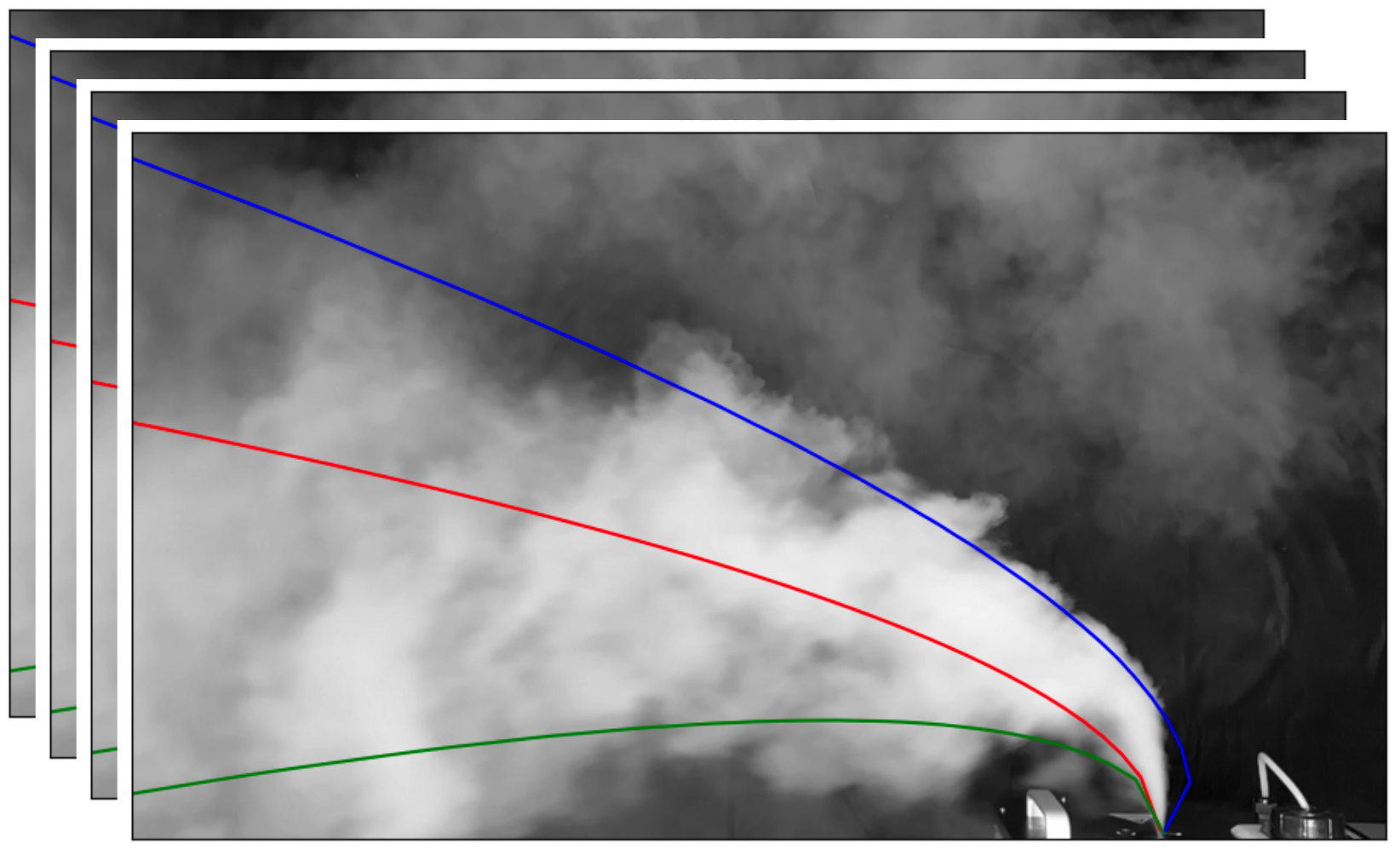}
        \caption{}
    \end{subfigure}

    \caption{A sketch of each step in this paper and the composite product.  (a) Cleaning the video and calculating the edge/center points.  (b) Testing different regressions to find the curve that best extrapolates center points for each frame.  (c) Modeling the spread of the edge points from the center path, and (d) the composite discovered centerline and edge dispersion, drawn on top of the original plume video}
    \label{fig:01}
\end{figure}

In this work, we introduce a data-driven framework that identifies a reduced-order analytical model of point-source plume dynamics, agnostic to plume type, directly from video data. Our framework operates in a two-step procedure: (i) time series extraction from video data (ii) and reduced-order model (ROM) discovery from the extracted time series data, described in detail in section \ref{sec:methods}. For the time series extraction, we obtain a best fit of the plume's mean path in each video frame by performing various regressions along the path of highest pixel density. Next, for each video frame, we fit a linearly-growing sinusoid to the plume edge. This process extracts the polynomial and sinusoidal cofficients over time that characterize the plume's average trajectory and spread. An overview the pipeline can be seen in Figure \ref{fig:01}.  In section \ref{sec:experiments}, we interrogate our framework' s ability to automatically extract dominant plume features from video data by extensively testing plume (i) point reduction, (ii) center-line (mean) curve fitting, and (iii) edge-line (variance) curve fitting. Section \ref{sec:discussion} describes important future work, including time-dependent modeling of the plume center, time-dependent dimension reduction, and experiments needed to establish the utility of the model in situ.\\

\section{Methods}
\label{sec:methods}
\subsection{Video Pre-processing}
  An unprocessed video of a plume that is recorded with ideal background conditions.\footnote{A solid black background is used when filming plume dynamics.}
  All frames are converted to gray scale, that is $n\times d$ arrays taking on values between $0$ and $1$ ({\color{black} or 0 to 255}). Two main steps are used in pre-processing prior to applying our model: background subtraction and Gaussian blurring. \\
  
  \noindent \textbf{Background subtraction.} A fixed background subtraction method is applied to isolate the plume, where the first $k_{\text{fixed\_subtraction}}$ frames of video (where no plumes are have formed) are used to create an average background image.
  The average image is subtracted against the remaining frames to create the isolated plume frames.  Once the plume dynamics have been isolated from background across selected frames, Gaussian blurring is applied. \\
  
  \noindent 
\textbf{Gaussian Blurring.} Two separately tuned Gaussian filters are applied---a temporal and spatial filter, respecitvely.  First, a Gaussian filter is applied across the time series of frames to add a time blur, reducing the high resolution of the plume dynamics. Second, a Gaussian filter is applied to each frame independently to coarse grain the image, adding spatial blur.

\subsection{Coarse-graining for Centerline and Edge modeling}

We denote the output of data pre-processing as $\boldsymbol{Z} = [\boldsymbol{z}(t_0), \hdots, \boldsymbol{z}(t_K)]$, where background subtraction and Gaussian filters have been applied. We extract a time series of second-order polynomial coefficients that model the center path of the plume for each frame. Additionally, we learn the parameters of a growing sinusoidal function that best characterizes the spread, or edge-model, of the plume. We theorize there exists a connection between the Kevin Helmholtz shear velocity and the sinusoidal frequency. 

\begin{figure}[htbp]
    \centering
    \includegraphics[width=\textwidth]{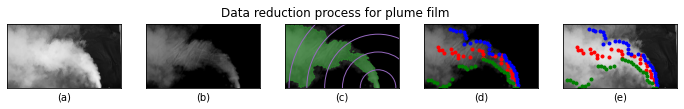}
    \caption{The plume point reduction step converts a frame of video into a scatter of edge and center points: (a) the raw frame image, (b) background subtraction, (c) contour selection, and finally (d) Along various ranges from origin, identify the max intensity (center) and intersection (edge) points with contours. (e) displays the final points on the original frame.}
    \label{fig:plume-steps}
\end{figure}

\bigskip
Each frame $\boldsymbol{z}(t_i)$ is converted to a reduced order model describing the center and edge paths in a three step procedure: (i) contour detection, (ii) a concentric circle search, and (iii) regression. For each array, $\boldsymbol{z}(t_i)$, image recognition techniques are used to identify the plume contour and subsequently search along concentric circles, centered at the leak source, to identify the path of highest density, where we denote raw pixel value as a proxy for plume density, and the edge paths, as seen in Fig. \ref{fig:plume-steps}. 

\begin{enumerate}[(i)]
    \item \label{pipeline: step 1}\textbf{Contour Detection.} We apply a binary threshold to identify the contours outlining the plume for each array $\boldsymbol{z}(t_i)$. Hyperpameter selection for the thresholding is done by using \texttt{opencv}'s Otsu's binarization. Optimal global threshold selection is performed by inspecting the image histogram for each array $\boldsymbol{z}(t_i)$. The $n$ largest contours, by area, are then selected for remainder of the pipeline. 
    We denote the identified plume at time $t_k$ as $\boldsymbol{\psi}_k$.
    \item \label{pipeline: step 2}\textbf{Concentric Circle Search.} A search is performed along a set of $\ell$ concentric circles, centered at the plume leak source, with incrementally growing radii of $r_i = r\cdot i$ for $i=1, \hdots, \ell$ where $r$ is some fixed positive value. Along each concentric circle, three values are attained: the largest pixel value that lies within the identified contour, $\boldsymbol{\psi}_k$, and the two intersection points of the concentric circle $r_i$ with the identified plume contour, $\boldsymbol{\psi}_k$. These denote the points for the center and edge paths respectively.
    \item \label{pipeline: step 3}\textbf{Regression.} Two regression techniques are implemented for the final steps for learning the centerline and edge plume paths. 
    \textit{Center Path.} Multiple second order polynomial regression are applied to the de-centered Cartesian coordinates of the centerline points identified in each frame from step \eqref{pipeline: step 2}. Which produces a timeseries of collected polynomial coefficients. 
    \textit{Edge Paths.} At each time, $t_k$, along each concentric circle with radii $r$ the Euclidean distances between the edge points and center point is attained. Giving set of distances for each radial value. We denote this process as \textit{flattening} the edge points. The flattened data is then bootstrapped and a series of growing sinusoid regression are applied via an Levenberg Marquardt optimization scheme. The mean set of values of the bootstrap bags are then selected for the final edge model.

\end{enumerate}

\section{Experiments}
\label{sec:experiments}
 We investigate each step of our plume data-reduction method in order to identify the best parameter choices.  To provide data for experiments, we filmed a smoke machine, where plumes were formed by vaporizing a water based triethylene glycol liquid, under three different intensities of crosswind, resulting in fifteen different videos.  Each video lasts for between twenty and one hundred seconds. Ten of the videos are designated training videos, and five are reserved to test conclusions drawn on the training videos, balanced between crosswind levels.  We conducted different experiments to learn the ideal parameters for fitting the model described in section \ref{sec:methods}: First, how best to reduce the greyscale frames to points along the center and edges of the plume; Secondly, how to fit a curve through the centerpoints that could extrapolate the direction of the plume, potentially off-frame.  Thirdly, how to identify spatiotemporal frequencies of the plume edge in hopes that these frequencies, associated with the Kelvin-Helmholtz instability, would reflect some physical parameters of the fluids.  The appendix contains additional information to reproduce the experiments. A pseduo code of the complete pipeline is shown in algorithm \ref{alg: full pipeline}.
\begin{figure}[htbp]
    \centering
    \includegraphics[width=0.8\textwidth]{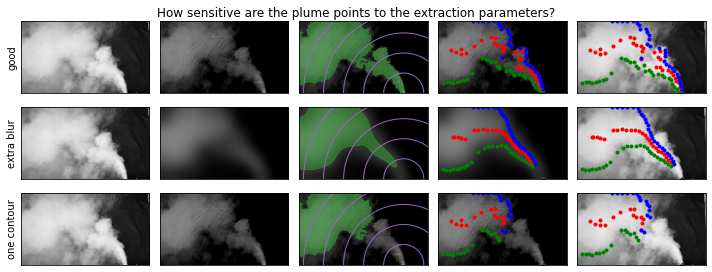}
    \caption{A visual comparison of the ideal parameters for our data set, along with selected less-than-ideal parameters.}
    \label{fig:step1_compare}
\end{figure}

\subsection{Plume point reduction}
At a macro scale, the first step is to transform video frames into scattered points representing the center and edges of the plumes.  We investigate parameters for space and time Gaussian smoothing, as well as more exotic OpenCV parameters for detecting contours and determining the intersections of the contours and concentric circles.  Because this step has no true value by which to evaluate goodness of fit, we manually review the same fifteen frames from each video's period of steady emission.  While it is possible to add additional smoothing via a stronger Gaussian blur or minimizing the number of contour points or concentric circles, we aimed to reserve smoothing explicitly for the subsequent, curve-fitting steps.  The goal of this step is merely to provide a good representation of the video.\\

We {\it a priori} identify the coordinates of the plume emission point in each video.  We compare the results of different parameterizations by visually evaluating the how well the plume points match the raw video.  Table \ref{tab:step1_compare} shows the different parameters we varied, and Figure \ref{fig:step1_compare} shows an example of a frame, comparing each step of plume point reduction against less-effective parameterizations.\\

The review of all the training frames suggested, and comparison of test frames confirmed, that a minor amount of spatial blurring and no time blurring were the best parameter selection.  In addition, since contour thresholding often rejected part of the plume, finding up to three contours improved the accuracy of plume point reduction. An increased amount of spatial blurring, when tested, provided little to no benefit in reconstruction. We conclude that results were not terribly sensitive to blurring. 
\begin{table}[!ht]
    \centering
    \begin{tabular}{c|c|c}
         Spatial Blur (px) & Time Blur (1 s/29.97)  & Number of contours \\
         \hline 
         0 & {\bf 0} & 1 \\
         {\bf 15} & 3 & 2 \\
         45 & 9 & {\bf 3} \\
         301 & 27 & 4 \\
    \end{tabular}
    \caption{The plume point reduction parameters compared across all trials.  We additionally tested out blurring more strongly along the direction of the plume, different versions of OpenCV parameters on contour thresholding (settling on ``OTSU") and concentric circle parameters (resulting in the saved defaults)}
    \label{tab:step1_compare}
\end{table}

\subsection{Center-line curve fitting}
The next step is to fit a curve to the center points of the plume.  With this in mind, we describe our coordinate origin as the plume emission point and compare four different curve-fitting methods: three that are a regression of $y=f(x)$ (linear, quadratic, and square-root), and one parametric quadratic fit that tries to identify $x = f(r), y = g(r)$.  It is worth noting that while the expression of the square root curve is the inverse of the quadratic curve, they are both fit on $y$-error, making the former nonconvex and bounded by a nonconvex domain.  To attempt to resolve these problems, we re-parameterize the square-root curve based upon the coordinates of the origin and the steepness of the curve and split the domain into four convex subdomains, fitting a constrained regression in each.  Additionally, while quadratic and square-root curves can be described as instances of a parametric quadratic curve, in implementation, the regression does not require the fitted curve to enforce $r^2 = x^2 + y^2$ (nor can it, while retaining quadratic terms). \\

\begin{figure}[htbp]
    \centering
    \includegraphics[width=1.0\linewidth]{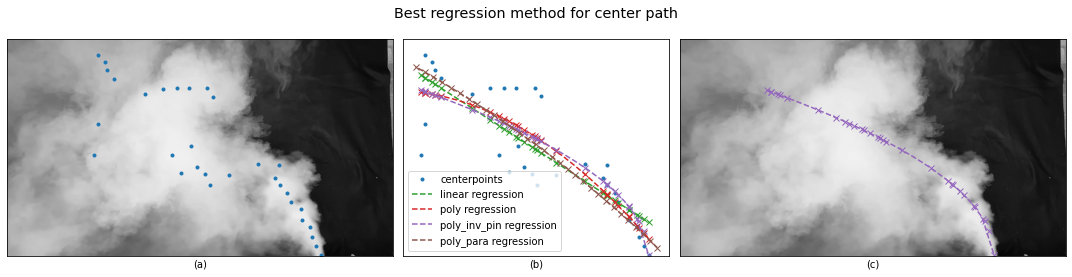}
    \caption{The experiment to find the best regression method for the plume path on a given frame: (a) The points of highest pixel intensity discovered by the previous step, overlayed on original frame. (b) A regression of each method against the centerpoints, up until the split between training and validation points.  (c) The regression with the lowest validation score, overlaid on the original frame.}
    \label{fig:step2a_example}
\end{figure}

We seek, in the situation where video data is close to a point emission source, to describe where the plume is drifting off-camera.  With this in mind, we split each frame into points \textit{near} the origin, used to fit a curve, and points \textit{away} from the origin, used to evaluate the curve, effectively creating a test regime for extrapolation. We compare the $l_2$ error of different curve fitting methods on the test points for each frame in order to determine the best method, as demonstrated in Figure \ref{fig:step2a_example}. \\

\begin{figure}[htbp]
    \centering
    \includegraphics[width=0.5\linewidth]{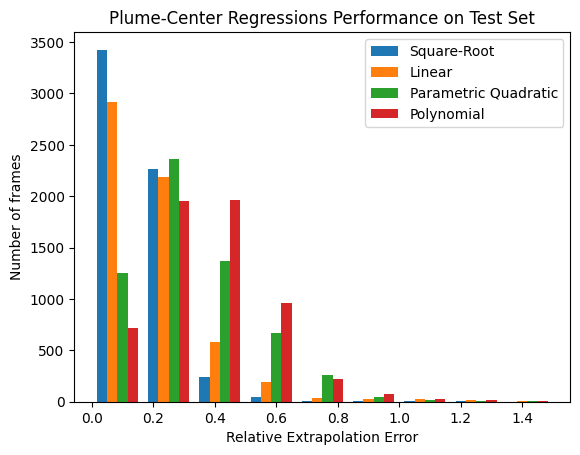}
    \caption{A histogram of the extrapolation errors for different regression methods across $\sim 6000$ frames of test data.  The square root regression performs the best, followed by linear, parametric quadratic, and then quadratic.}
    \label{fig:step2a_compare}
\end{figure}

From initial review of the training data, we hypothesized the square-root curve would be the best curve-fitting method based on visual inspection. We evaluated the results on the test dataset that comprises 6000 individual frames.  With information on the best fitting method for each frame, we concluded that the square-root curve was the best fitting method overall at the 99\% confidence level, with p-values near machine precision (see Table \ref{tab:center-pvals}).  However, the magnitude of the advantage over a linear regression was often small; relative error of extrapolation for each method is demonstrated in Figure \ref{fig:step2a_compare}.  The advantage was most significant in low-wind datasets, where the plume arcs throughout the frame, and mostly disappeared in high-crosswind datasets, where the force from high-crosswind would rapidly dominate the initial point-source ejection force of the plume.\\

The results showed that the square-root curve was effective at extrapolating the direction of the fitted points with less than 20\% relative error in most frames.  This is even more substantial in context of how the fitted points were created in the previous step: the assumption that points of highest pixel intensity represent the greatest concentration of the plume, and that the points of greatest concentration best described the path of the plume.  If, on the contrary, the plume's path at a particular radius is better described by its center of mass, or by the midpoint of the identified contours, the data fed into this experiment would naturally be smoother. \\

Nevertheless, the method and the experiments could be improved in a few ways.  The choice of evaluating methods based upon extrapolation may result in worse models when trying to understand the physics before the dissipative regime.  Changing the metric from extrapolation to interpolation would also unlock fitting with a spline curve, which allows parametric paths.  Additionally, one could replace the parametric curve with a function $\theta(r)$ in polar coordinates.  This would also make it easier to enforce the constraint that the fit curve crosses the origin, i.e. the plume emission point.  Currently, only the square-root curve regression incorporates knowledge about the origin or rather, about the extremal points and the direction of the plume---a better comparison would provide similar information to the alternate regression methods. \\

\begin{table}[!ht]
    \centering
    \begin{tabular}{c|c|c|c|c}
         & Root curve & Linear & Parametric-Quadratic & Quadratic \\
         \hline
         Root curve & N/A & 0.0 & 0.0 & 0.0 \\
         \hline
         Linear & 1.0 & N/A & 0.0 & 0.0\\
         \hline
         Parametric-Quadratic & 1.0 & 1.0 & N/A & 0.0 \\
         \hline
         Quadratic & 1.0 & 1.0 & 1.0 & N/A
    \end{tabular}
    \caption{P-Values for one-direction T-Test that method on left is more often a better curve-fitting method for extrapolation than method across top}
    \label{tab:center-pvals}
\end{table}

\subsection{Edge-line curve fitting}
The goal is to understand plume spread with accurate edge modeling and capture the frequencies connected to Kelvin-Helmholtz instability. After the initial laminar emission, the plume appears to exhibit instability, with the edge experiencing rapid irregularities, before ultimately moving to a dissipative regime off-camera. This could be characterized by Kelvin-Helmholtz instability between the velocity different between two fluids: namely the plume vapor and surrounding air. The frequency of oscillations would relate to the mass flow rate of the plume emission to which we hope to identify the frequency directly form video data. \\

We opt to model the edge spread as a distance from the mean path of the plume trajectory. That is, before any regression, the edge points identified along the contour of the plume, $\psi_i$, from the step of concentric circles, are converted from Cartesian coordinates $(x,y)$ to the ordered pair $(t,r,d)$--where $d$ is the distance from the edge point to the corresponding center point identified along the concentric circle with radii $r$, at time $t$. For any given video this is done across all edge points from all frames. This ``\textit{flattening}'' process transforms the data into a form more easily interpretable to apply regression. \\

\begin{figure}[htbp]
    \centering
    \includegraphics[width=1.0\linewidth]{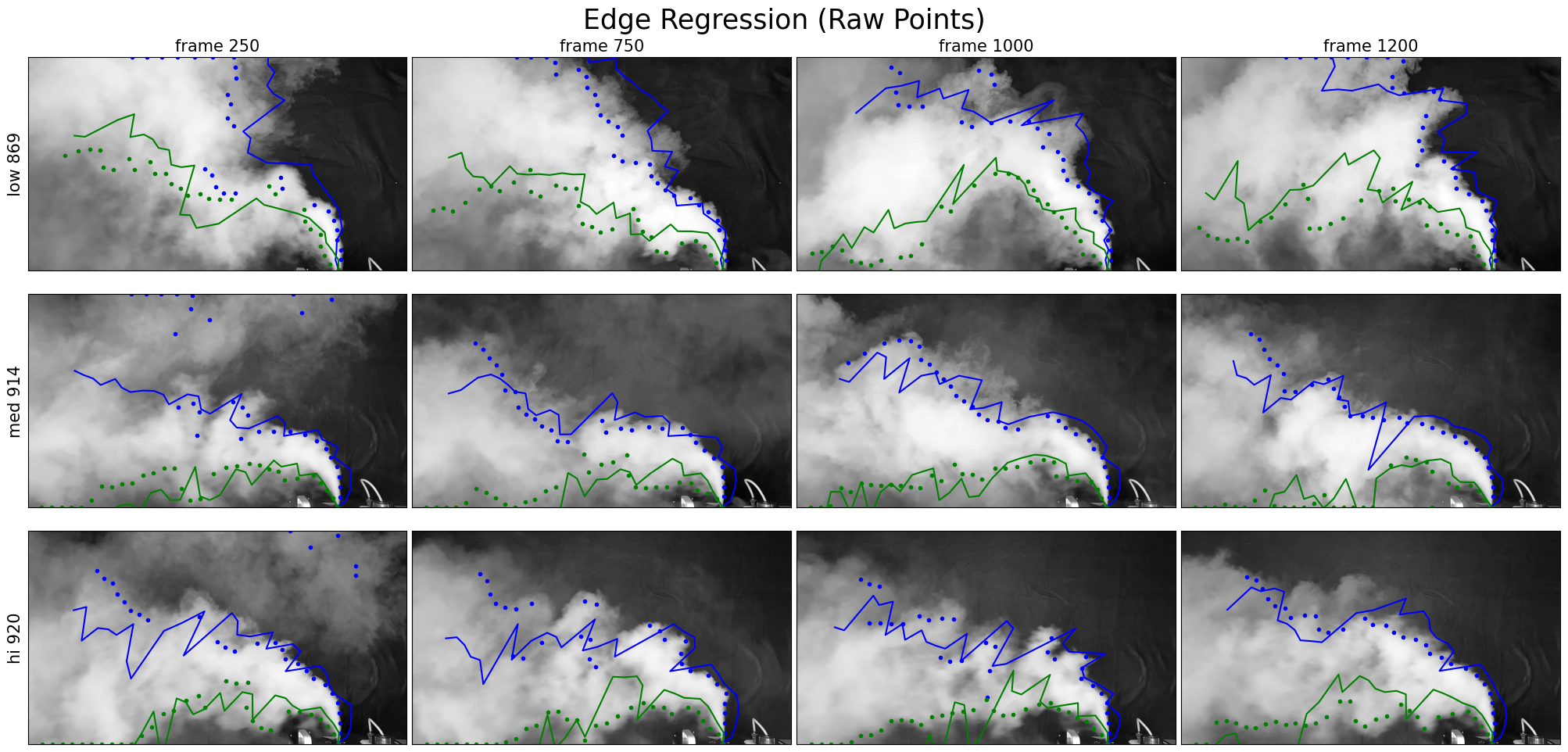}
    \caption{Edge regression path shifted by true center points detected. }
    \label{fig:step2b_example}
\end{figure}

Due to the oscillatory nature observed in the raw video data, we fit a growing sinusoid function of the form 
\begin{equation}
    d(r,t) \triangleq A \sin{(\omega r - \gamma t + B)} + C r + D
    \label{eqn: edge func}
\end{equation}
to the flattened data, where $(A,\omega, \gamma, B, C, D)$ are to be discovered. For all videos a bootstrap method was used to reduce data variability. Approximately $80\%$ of the data was randomly selected for training. Upon which, sampling with replacement was done to create $2000$ to bags to fit equation \eqref{eqn: edge func} too. Fitting is done by solving the unconstrained optimization problem via the Levenberg-Marquardt algorithm, to which if no optimal parameters are found within a given number of iterations, the trail is rejected and disregarded. The mean parameters from the non rejected trials are selected for the final model output. The learned function is mapped back onto the original frames by a processing of \textit{unflattenning}. Where points calculated from $\eqref{eqn: edge func}$ are shifted by their polar angles against either the true center points or the learned center path regression. An example of points mapped back onto the original frames, where shifts were applied against the true center points can be seen in figure \ref{fig:step2b_example}. \\ 

\begin{figure}[htbp]
    \centering
        \includegraphics[width=1.0\linewidth]{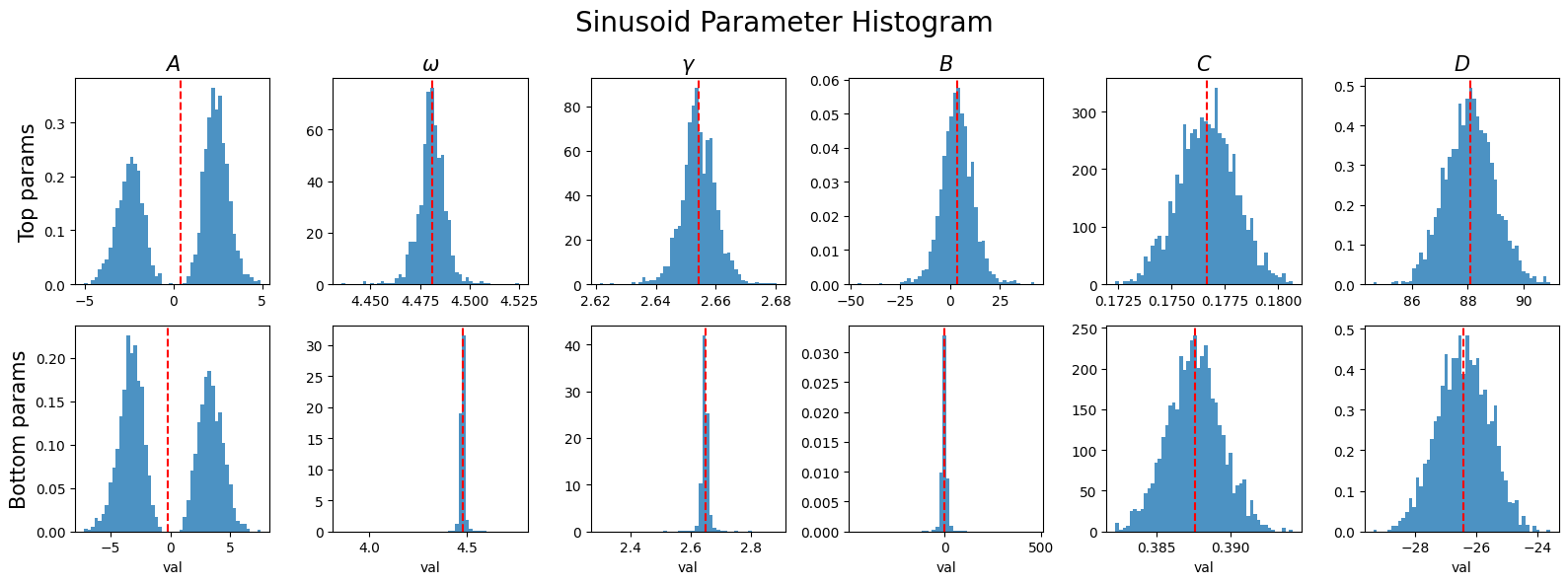}
    \caption{Histogram of amplitude and frequencies for video $920$ across $2000$ bootstrap bags. Top row denotes learned parameters for top sinusoid fit and bottom row is for the bottom edge of plume. The redline denotes the mean selected parameters used for the final model. }
    \label{fig:step2b_hist}
\end{figure}

A distribution of learned parameters can been seen in figure \ref{fig:step2b_hist}. Where we note the observations (occurring frequently across all videos) namely the tight distribution for the ``bottom'' set of parameters and the bimodal behavior of the amplitudes, $A$. In general both the train and validation accuracies, shown in table \ref{table:video_acc}, tended to also be higher on the ``bottom'' fit of frames. This is likely a computational artifact of the plume formation progressing off screen at the bottom, causing edge points to be identified along the boundaries of the video. Leading to a simple function to be fit. \\

Despite the narrow range of frequencies, the amplitude appears bimodal: positive and negative modes offset each other, resulting in a mean near zero.  Such bimodal behavior, exhibited in figure \ref{fig:step2b_hist}, could be an artifact of the nonconvex optimization problem, but could just as well indicate a nonlinear phase shift. In order to investigate further, an alternative approach to frequency identification would be a 2D Fourier transform.  Because our data is ragged, such a Fourier transform would need to be formulated as a matrix completion problem, but would be convex and could provide additional insights to the instability based upon a range of frequencies. \\

\begin{table}[!ht]
\centering
\begin{tabular}{|c|cc|cc|}
\hline
\textbf{Video ID} & \multicolumn{2}{c|}{\textbf{Train Accuracy}} & 
\multicolumn{2}{c|}{\textbf{Validation Accuracy}} \\ \hline \hline
                  & \textbf{Top}          & \textbf{Bot}           & \textbf{Top}          & \textbf{Bot}          \\ \hline
low 862           & 0.6166                & 0.3847                 & 0.6097                & 0.3910                \\ \hline
low 865           & 0.4681                & 0.5089                 & 0.4689                & 0.5106                \\ \hline
low 867           & 0.4005                & 0.4957                 & 0.3982                & 0.5073                \\ \hline
low 869           & 0.4063                & 0.4985                 & 0.4083                & 0.5026                \\ \hline
low 913           & 0.5581                & 0.6544                 & 0.5563                & 0.6501                \\ \hline
med 871           & 0.4835                & 0.4937                 & 0.4868                & 0.4936                \\ \hline
med 914           & 0.4610                & 0.6150                 & 0.4562                & 0.6151                \\ \hline
med 916           & 0.4279                & 0.4821                 & 0.4280                & 0.4801                \\ \hline
hi 919            & 0.4180                & 0.4992                 & 0.4172                & 0.5059                \\ \hline
hi 920            & 0.5006                & 0.5364                 & 0.4939                & 0.5411                \\ \hline
\end{tabular}
\caption{Train and Validation Accuracy (Top/Bottom)}
\label{table:video_acc}
\end{table}

\section{Discussion}
\label{sec:discussion}
\begin{figure}[htbp]
    \centering
    \includegraphics[width=1.0\linewidth]{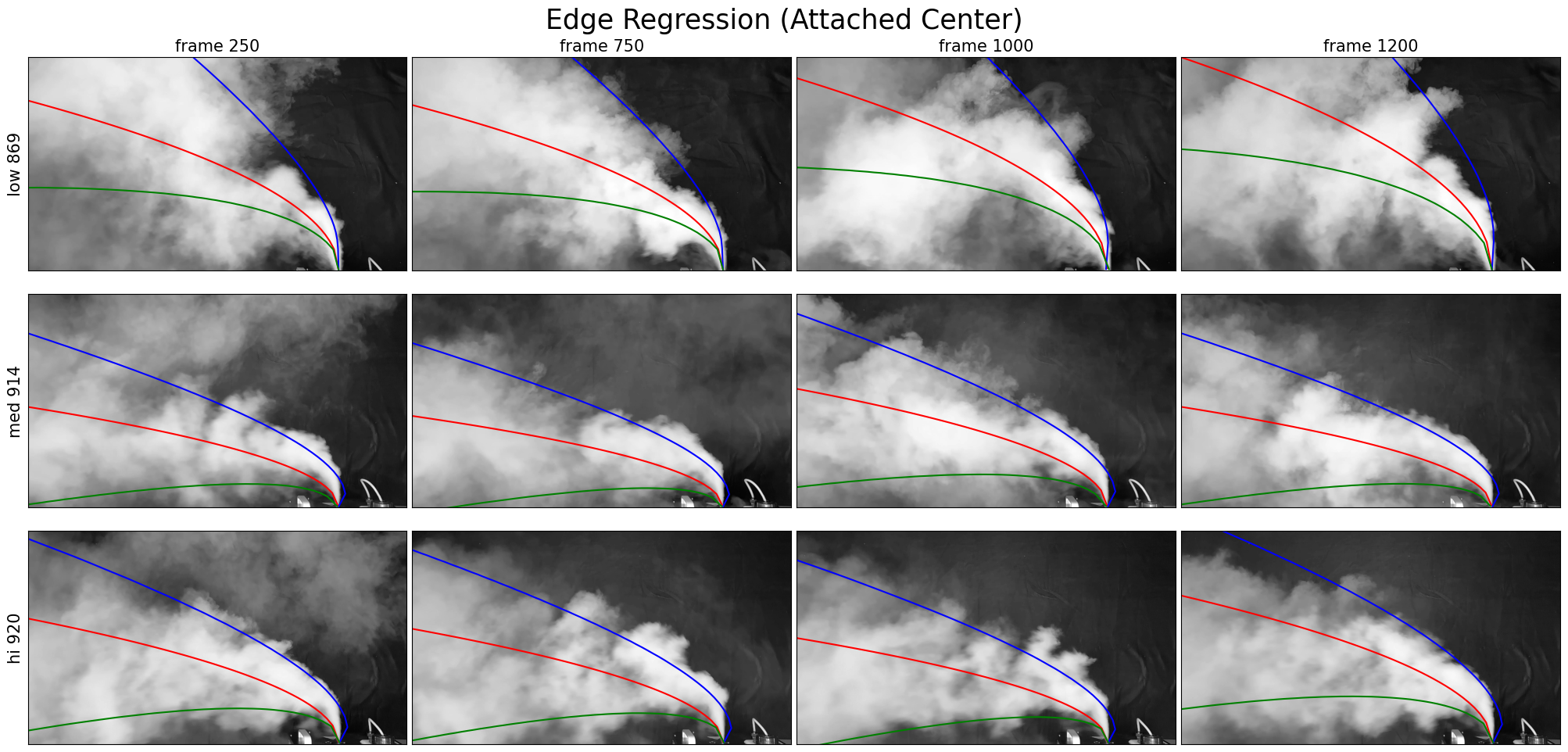}
    \caption{The complete model applied to a video. Edge model is shown attached to the regressed center, rather than the raw center points.}
    \label{fig:all-together}
\end{figure}
The result of the complete pipeline is demonstrated in Figure \ref{fig:all-together}. Experiments have borne out the decision to treat the plume as following a square-root curve, which maintains low relative error in extrapolation.  Alternate curve-fitting performed poorer in extrapolation, though promising variations remain unexplored.  For modeling the edge of the plume, we found that a linear relationship, constant in time, does reasonably well to describe the edge of the plume as it drifts away from the source.  While regression found substantial sinusoidal behavior, more work needs to be done to identify frequencies associated with the Kelvin-Helmholtz instability, and from there determine plume emission rate. \\

\begin{algorithm}
\caption{Video to Reduced Order Model}
\begin{algorithmic}[1]
    \State \textbf{Clean video}
    \State \quad 1a. Background subtract
    \State \quad 1b. Apply Gaussian blur
    
    \State \textbf{Calculate Plume Points}
    \State \quad 2a. Compute contours using Otsu's method
    \For {each range ring from plume origin to edge of image}
        \State 2bi. Of all contour pixels crossed by the Concentric Circle (CC), select the pixel with max intensity as center
        \State 2bii. Find intersections of contour and CC
        \State 2biii. Perform barycentric interpolation of intersections onto the range ring
        \State 2biv. Choose the furthest clockwise (CW) and counterclockwise (CCW) interpolants as edge points
    \EndFor

    \For {each frame}
        \State Regress the height of the plume center as a square-root function of horizontal distance
    \EndFor

    \State Regress edge points for each frame
\end{algorithmic}
\label{alg: full pipeline}
\end{algorithm}

This effort fits within the ambit of ``Go Pro physics", which seeks to conduct investigations on raw video of dynamical phenomena.  In the real world, however, conditions will be less than ideal.  Issues such as camera jitter or low plume contrast would degrade the video, and its unclear how the difficulty would scale with these introduced factors. In particular, we aimed to remove as much Gaussian blur as possible, as additional smoothing in the first step would buoy regression metrics, potentially obscuring problems in our method.  When using noisier data from the field, more blurring may be needed. \\

A more robust dataset would include not just more videos, but from more angles and with a wider aspect in order to capture higher Reynolds number behavior.  The experiments could also be improved, as mentioned in their respective sections.  But more ambitiously, once we can get useful frequencies from the we could simulate a plume using an accurate CFD engine such as OpenFoam\footnote{We attempted with Blender, but it only simulates in the purely dissipative regime.}  This would allow us, knowing the physics parameters a priori, to evaluate how well the discovered frequencies reflect the Kelvin-Helmholtz frequencies and thus the mass flow rate.  In doing so, it may help to cluster the plume points in space and time in order to separate the laminar flow, unstable flow, and turbulent dissipation, e.g. using Spatio-temporal K-means \citep{Dorabiala2022SpatiotemporalK}.\\

More ambitiously, it may be possible to get a more robust model from using time-series data reduction models such as Dynamic Mode Decomposition \citep{kutz2016dynamic} or Sparse Identification of Nonlinear Dynamics \citep{Brunton2016} directly on the raw video.  These methods may also be able to directly recover the oscillatory frequency or fluid parameters of interest. \\

\section{Conclusion}
\label{sec:conclusion}
We have produced a reduced-order model for a point-source plume that, with enough crosswind, can describe the behavior of a plume on video.  The model captures the dominant features of nonlinear plume dynamics using a low dimensional representation that, in contrast to Galerkin models, extrapolates beyond the spatial bounds of the frame.  Against a handful of lab-condition but varied datasets, we have refined the parametrization and calibrated its effectiveness.  We have opened the door to questions about using this reduced-order model to understand important physical properties of the plume: how much is being emitted in the first place, and how will it dissipate.  These questions could, in the long run assist the management of industrial accidents, monitoring of pollution, and understanding natural phenomena such as volcanic and geothermal emissions.\\

We propose a natural extension to parameter initialization for Gaussian plume modeling, addressing a key challenge in this well-established technique—determining the initialization that corresponds to a desired plume formulation. Our approach integrates a data-driven paradigm to bridge video data with the commonly used first-principles model for plume dynamics, capturing the transition from source-dominated forces to turbulent diffusion and advection. Additionally, we have developed the open-source Python packages \texttt{rom-plumes} and \texttt{plumex} to facilitate experiment creation and replication.

\section*{Acknowledgements}
We would like to thank Laurel Doyle, Applied Mathematics at University of Washington for her early experiments and investigative work into plume modeling in contribution to this paper.  

\section*{Appendix}
Code for the experiments are split between two Python packages: rom-plumes ~\citep{plumes-github} and rom-plumex ~\citep{plumex-github}. The former includes all functionality a user needs to apply the methods in this paper to their own data.  The latter includes the the experiments in this paper, experimental configuration, as well as the code to regenerate the figures.   Each experiment is a callable, run through the command line program mitosis ~\citep{Stevens-Haas_mitosis_2024}.  Cloning the rom-plumex repository and editably installing it will install all the required dependencies for the experiments.  To exactly reproduce the results of the experiments, download \cite{zenodo_data} and unzip as a folder plume\_videos in the repository.  We have uploaded the mitosis artifacts of our experiment trials to \cite{zenodo_experiments}, which includes an environment file (requirements.txt) and a source file (source.py), among other plot artifacts, logs, configuration, and data.  Once the plume\_videos folder is in the repository, installing a particular environment file and running the source file should produce the visual results in the experiment.html file and the python objects in results\_0.dill, results\_1.dill, etc. where each file refers to a different step of the experiment.  To re-run the experiments with different parameters, see the documentation for mitosis. While different experiments were run on different commits of rom-plumex, the final commit as of publication is tagged 0.2
\bibliography{main}

\end{document}